# Remotely induced magnetism in a normal metal using a superconducting spin-valve

M.G. Flokstra[1]*, N. Satchell[2], J. Kim[2], G. Burnell[2], P.J. Curran[3], S.J. Bending[3], J.F.K. Cooper[4], C.J. Kinane[4], S. Langridge[4], A. Isidori[5], N. Pugach[5,6], M. Eschrig[5], H. Luetkens[7], A. Suter[7], T. Prokscha[7], S.L. Lee[1]

[1]School of Physics and Astronomy, SUPA, University of St. Andrews, St. Andrews KY16 9SS, United Kingdom.

[2]School of Physics and Astronomy, University of Leeds, Leeds LS2 9JT, United Kingdom.

[3]Department of Physics, University of Bath, Claverton Down, Bath BA2 7AY, United Kingdom.

[4]ISIS, Rutherford Appleton Laboratory, Oxfordshire OX11 0QX, United Kingdom.

[5]SEPnet and Hubbard Theory Consortium, Department of Physics, Royal Holloway, University of London Egham, Surrey TW20 0EX, United Kingdom.

[6]Skobeltsyn Institute of Nuclear Physics Lomonosov Moscow State University (SYNP MSU), Leninskie Gory, Moscow 119991, Russia.

[7]Labor für Myonspinspektroskopie, Paul Scherrer Institut, CH-5232 Villigen PSI, Switzerland.

*Correspondence to: mgf@st-andrews.ac.uk.

**Abstract**

Superconducting spintronics has emerged in the last decade as a promising new field that seeks to open a new dimension for nanoelectronics by utilizing the internal spin structure of the superconducting Cooper pair as a new degree of freedom. Its basic building blocks are spin-triplet Cooper pairs with equally aligned spins, which are promoted by proximity of a conventional superconductor to a ferromagnetic material with inhomogeneous macroscopic magnetization. Using low-energy muon spin rotation experiments, we find an entirely unexpected novel effect: the appearance of a magnetization in a thin layer of a non-magnetic metal (gold), separated from a ferromagnetic double layer by a 50 nm thick superconducting layer of Nb. The effect can be controlled by either temperature or by using a magnetic field to control the state of the remote ferromagnetic elements and may act as a basic building block for a new generation of quantum interference devices based on the spin of a Cooper pair.

**Main Text**

The ability to manipulate the spin degree of freedom of charge carriers is key to realizing future spin-based electronics. Integrating superconductors into spintronic devices can greatly enhance performance[1] and allows the transport of spin over long distances without the dissipation of heat[2]. In order to achieve the alignment of electron spins ferromagnetic materials are used. Superconductivity and ferromagnetism are, however, antagonistic states of matter, and the interplay between these two states results in the conversion of conventional spin singlet into spin triplet pair correlations[3]. Whereas spin singlet pairs have spin angular momentum $S = 0$, spin triplet pairs have $S = 1$ with three possible spin projections $s_z = -1, 0, +1$. The realization of such spin-triplet pairs in mesoscopic systems containing interfaces between superconducting (S) and ferromagnetic (F) layers has attracted much interest from both the theoretical and experimental communities. Interaction of spin-singlet superconductivity with collinear ferromagnetism leads to oscillations and suppression of the pair correlation at a short distance $\xi_f$ due to the exchange magnetic field in the ferromagnet, which tends to align the spins of electrons parallel[4-7]. However, in order to create longer-range penetration of spin-triplet superconductivity into the ferromagnet, interaction with a non-collinear magnetism is required[8-10] motivating the discovery of superconducting currents through ferromagnetic metals over distances far longer than the singlet penetration length $\xi_f$[11-13]. These long-range triplet components (LRTC) have parallel spin projections ($s_z = \pm 1$), and are not suppressed by the exchange field. Theory predicts that the conversion into spin triplet pairs should also give rise to an induced magnetic moment in the superconductor, decaying away from the interface[14-16], often called the inverse or magnetic proximity effect. For diffusive systems this induced magnetic moment is predicted to be negative (opposite to the magnetization of itinerant electrons in the adjacent F layer) and accompanied by a small decrease of magnetization of this F layer on the scale of the ferromagnetic coherence length $\xi_f$. There are a small number of reports with observations that are attributed to this effect[17-19] though none use a measurement technique that has the required spatial sensitivity to uniquely determine this. A further report involving low-energy muon spin rotation (LE-µSR) measurements, a technique possessing the required spatial sensitivity to determine the location of the moment, found contradictory evidence[20]. The moment was found not to penetrate into the S layer over the expected distance of a coherence length, but rather it existed over a very much shorter length scale, indicating a rather different interfacial mechanism at play in that system and possibly also in related works.

Here we report results obtained by high precision LE-µSR that are in conflict with the current theoretical predictions, and which yield instead a very surprising, hitherto unknown effect. We find a switchable magnetic moment to be induced remotely from the superconductor-ferromagnet interface, at a nonmagnetic superconductor-normal metal interface about 150 atomic layers away from the ferromagnet. The moment appears, however, not inside the S layer, but in an adjacent normal metal (N) layer. It first appears at the onset of superconductivity and increases as the temperature is lowered. This remote induced magnetic moment also exhibits a spin-valve effect: a significant change in magnitude (~20 times) depending on the mutual orientation of magnetization in the F layers in the NSFF multilayered structure. The effect almost disappears when switching the spin-valve into a collinear state of the F layers' magnetization, when LRTC are absent. This shows that LRTC in the ferromagnetic regions are a crucial ingredient contributing to the effect.

For our experiments we use superconducting spin-valve structures Au(x) / Nb(50) / Co(2.4) / Nb(3) / Co(1.2) / IrMn(4) / Co(3) / Ta(7.5) / Si-substrate with numbers indicating the layer thicknesses in nm and x = 5 or 70. They consist of an S/F interface with an additional N layer atop the S, as well as a second F layer separated from the first by a thin normal metal spacer (n) creating a NSFnF device, shown schematically in Fig.1 (see supplementary information for more details of our spin valves). In our devices the exchange field of the outer F layer (Co(1.2)), can be pinned magnetically, by using an anti-ferromagnet (IrMn), while retaining easy manipulation of the other F layer (Co(2.4)). This enables us to control the angle between the two F magnetizations and thus to explore the inverse proximity effect in both the orthogonal configuration as well as the collinear configuration. In other words to examine the (possible) induction of magnetic moments when the LRTC are present (noncollinear configuration) and compare it with the case where they are absent (collinear configuration). A dependence of $T_c$ on the magnetic configuration in such structures has been proposed[21] and measured[22-24]. For the case of a strongly spin-polarised ferromagnet, due to appearance of the new LRTC channel for drainage of Cooper pairs from the S to the F layers, the change of $T_c$ between the collinear and perpendicular configuration may be much more pronounced than between parallel and antiparallel alignment[24].

To study the flux profile $B(y)$ as a function of depth $y$ in our superconducting spin-valves we use LE-μSR at low temperatures (3-10 K). During a muon experiment, low energy spin-1/2 muons (~4-26 keV) are implanted into the sample at normal incidence to the sample surface. The actual implantation profile depends on the muon energy (see Fig.1) and can be accurately calculated using Monte Carlo simulations[25]. Once implanted, the muon spin starts to precess around the local field direction with a frequency that is proportional to the local field strength, before it eventually decays and emits a positron preferentially along its momentary muon spin direction, allowing the time evolution of the muon spin to be monitored. LE-μSR is an exquisitely sensitive technique with which to determine the local flux density with a spatial resolution better than the coherence lengths involved. We compare the flux profile $B(y)$ obtained above and below the superconducting transition temperature in order to study the remote proximity effect and to demonstrate its connection to superconductivity.

A typical approach to fitting the muon data for a particular implantation energy is to use standard model functions characterized by the average flux $\langle B \rangle$ across that stopping profile[25]. Repeating this for a range of implantation energies, each corresponding to a different average depth $\langle y \rangle$ into the sample, provides a good indication of the spatial dependence of $\langle B \rangle(\langle y \rangle)$. A more sophisticated approach to modelling involves combining information from all implantation energies and fitting simultaneously to a common $B(y)$ describing the actual flux profile across the sample depth[24] while taking into account the full stopping profiles of the muons. A series of measurements were made, varying the implantation energy at fixed temperatures, allowing both types of analysis to be performed (see supplementary information for more details on the data treatment).

The main results of the analysis of our LE-µSR data are presented in Fig.2A. The induced magnetic profile $B(y)$ is presented as a function of position for orthogonal and collinear arrangements, determined both above ($T = 10$ K) and below ($T = 3$ K) the superconducting transition temperature ($T_c \sim 7.5$ K). Above $T_c$ the magnetic profile obtained, for both arrangements, is approximately constant at the external field of 150 G. However, upon cooling to below $T_c$ a sudden appearance of a magnetic induction in the Au layer is obtained for the orthogonal arrangement, which almost completely disappears in the collinear arrangement (in our experiments we probe the parallel aligned collinear state). Additionally, inside the superconductor no observable change is detected for either magnetic state, thus indicating that the Meissner screening is unobservably small. This is consistent with earlier findings[20], reflecting both the thinness of the superconducting layer and the strong suppression of the superconducting order parameter by proximity to ferromagnetism. Fig.2B shows a comparison between both types of modelling, where the $\langle B \rangle(\langle y \rangle)$ obtained for each individual dataset (square symbols) are compared to the calculated values from the results shown in Fig.2A (solid lines). The generally good agreement shows the obtained $B(y)$ is indeed a good representation of the actual magnetic profile (see supplementary information for more details of alternative fitting functions). When comparing the behavior in the superconducting and normal states, the results can be summarized as follows. 1) A magnetization is induced in the normal metal with a sign opposite to the magnetization direction of the free F-layer (since it subtracts from the applied field of 150 G), which decays away towards the surface of the sample on a scale ~20 nm. 2) This effect is clearly visible in the orthogonal arrangement but diminishes (by a factor of 20) for the collinear arrangement. 3) Unexpectedly, no induced magnetization is observable in the superconducting layer. All these facts are inconsistent with the theory[14-16] of the inverse (magnetic) proximity effect.

The temperature dependence of this effect which disappears above $T_c$ shows a clear correlation with the onset of superconductivity (see Fig.2C). This demonstrates that the S layer, itself not being spin-polarized, nevertheless provides this nonlocal magnetic effect. To further examine this absence of induced moment in the superconductor we measure a sample with a much thinner (5 nm) normal metal cap but otherwise identical to the sample from Fig.2A, in the orthogonal arrangement. This allows the superconductor layer to be probed directly without mixing in a large contribution from the N cap. No difference in the field profiles with temperature is observed for muon energies that probe the sample up to the interface with the F layer (see Fig.3).

Nevertheless a small contribution of an additional positive magnetization (along the external magnetic field) was detected at the highest muon energy where muons also stop in the FnF region which thus contributes to the signal.

Current theories do not account for our observed effect, and two main facts require explanation: 1) the remote magnetization provided by superconductivity of the interlayer, and 2) its dependence on the mutual orientation of the F layers magnetization. Here we propose potential mechanisms to understand these results (see supplementary information for further details). The first question to address is how a thick superconducting layer, itself not being magnetized, may provide the transfer of magnetization (or spin polarization) from the FnF region to the N layer. We envisage two possibilities: the first being spin transfer by crossed Andreev reflection (CAR) and elastic co-tunneling (EC)[26] and the second being spin transfer by pure spin currents. The former involves spin-singlet pairs either being formed from electrons originating from the interfaces at opposite sides of the S layer (CAR) or being used to effectively transfer an electron from one of the interfaces to the other interface (EC). The alternative involves flows of spin-triplet pairs (and is thus a direct consequence of having LRTC in the system) where a net flow of spin-up electron pairs moving from one side of the S layer to the other side is cancelled by an opposing flow of spin-down electron pairs. These mechanisms are illustrated in Fig.4.

The second question to address is the observed spin-valve effect: the disappearance of the remote magnetization together with the LRTC at the collinear magnetic configuration. To transfer the observed negative magnetization into the N layer by the CAR or EC mechanism, some negative spin accumulation must exist near the S/F interface. Spin accumulation itself appears as a result of spin current decay[27] (it could also be ascribed to the inverse proximity effect[14] but since that wouldn't result in spin-valve behavior, we exclude it as a candidate mechanism). It was shown that spin currents, both normal[28] and superconducting[29-30], appear in FnF spin-valves with noncollinear spin alignment (where LRTC are present), even in an unbiased structure, but disappear in the collinear geometry (where LRTC are absent). Thus spontaneous spin currents in the FnF region can lead to spin accumulation in the N layer by CAR and EC processes.

In conclusion, our results demonstrate that the induced magnetization is directly connected to the presence of the LRTC, can be experimentally controlled by switching magnetization or temperature, and appears remotely (> 50 nm) from the interfaces that generate these spin triplet correlations. This motivates a re-evaluation of the theoretical mechanism leading to the observation of a measureable magnetization generated by spin triplets. In such spin valves large-scale superconducting correlations together with spin-singlet – spin-triplet conversion provide this nonlocal magnetic effect. Our experiments demonstrate that a significant magnetization can be generated in a normal metal due to its proximity to a superconducting source, attributable to a spin triplet spin current that is controlled by a magnetic switch. This involves no charge current and no driving voltage. The magnetization in the N metal can only arise due to a net accumulation of excess spin. The separation of charge and spin currents and the induction of spin polarized electrons are building blocks of spintronic devices and our experiments provide a novel mechanism by which such devices might be realized.

# Methods

**Sample fabrication.** Samples were prepared by dc magnetron sputtering at a base pressure of $10^{-8}$ mbar. Layers were grown *in situ* on Si(100) substrates at ambient temperature at a typical growth rate of 0.2 nms$^{-1}$. The layout of our spin-valves is Au(x) / Nb(50) / Co(2.4) / Nb(3) / Co(1.2) / IrMn(4) / Co(3) / Ta(7.5) / Si-substrate with numbers giving the layer thickness in nm and x = 5 or 70. Growth was performed in the presence of a homogeneous magnetic field at the sample to establish the magnetic pinning of the Co layers adjacent to the IrMn, where the bottom Co layer is needed to set the initial direction for the IrMn to be pinned. The Ta buffer layer is to improve growth quality and the Au capping layer has a dual purpose. It protects the sample from oxidation and, depending on its thickness, allows the muons to either probe the Nb layer directly (5 nm Au cap) or to probe the observed proximity effect in the Au layer (70 nm Au cap).

**LE-µSR measurements**. The low energy muon spin rotation (LE-µSR) experiments have been carried out at the µE4/low energy muon (LEM) beamline[31] of the Swiss Muon Source as described in SI 3.1. For all measurements the applied field was oriented in the sample plane, either perpendicular to the pinning direction (orthogonal arrangement) or aligned with the pinning direction (collinear arrangement). The field used to attain saturation of the free Co layer was 150G. Temperature scans at fixed muon implantation energy were performed over a temperature range of 3 to 20K, while energy scans were made both above $T_c$ as well as below $T_c$. Typically 2 to 6 million muon decay events were counted for each muon experiment.

# Acknowledgements

We acknowledge the support of the EPSRC through Grants No. EP/J01060X, No. EP/J010626/1, No. EP/J010650/1, No. EP/J010634/1, and No. EP/J010618/1, support of a studentship supported by JEOL Europe and the ISIS Neutron and Muon Source, and the support of the RFBR via awards No. 13-02-01452-a, and No. 14-02-90018 Bel-a. All muon experiments were undertaken courtesy of the Paul Scherer Institute.

# Author contributions

J.K. and G.B developed the samples; M.G.F., S.L.L., N.S., J.F.K.C, H.L. and T.P. performed the muon measurements where H.L., A.S. and T.P. provided the beamline support; M.G.F., S.L.L., N.S., J.F.K.C., P.J.C., S.J.B., C.J.K. and S.L. performed various support and characterization measurements; A.I., N.P. and M.E. provided theoretical interpretation of the data and helped writing the paper; G.B. and M.E. helped designing the study; M.G.F. and S.L.L. designed the study, analysed data and wrote the paper. All authors discussed the results and commented on the manuscript.

# Competing financial interests

The authors declare no competing financial interests.

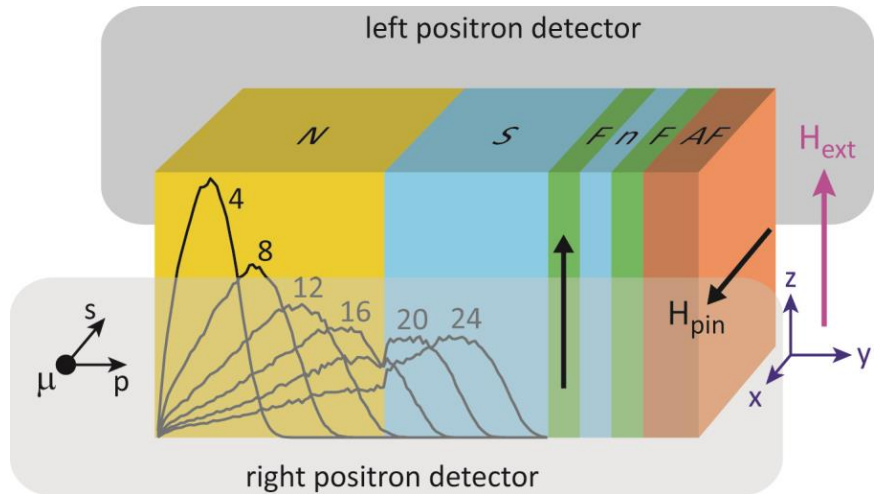

**Fig. 1**. **Sample architecture and experimental arrangement.** Schematic of the sample architecture (NSFnF), centered between the positron detectors within a homogeneous applied field ($H_{ext}$) along the z-direction. The momentum (**p**) of the incoming muon (μ) is normal to the sample plane (along the y-direction) and its initial spin (**s**) points towards the left positron detector. The direction of the exchange field of the (free) F layer closest to the S layer is saturated along the applied field direction, while the second (pinned) F layer is always directed along the pinning direction from the anti-ferromagnet ($H_{pin}$). The sample orientations used were either with $H_{pin}$ aligned with $H_{ext}$ (collinear arrangement) or perpendicular to it (orthogonal arrangement). Muon stopping profiles are overlayed on the front face of the sample to indicate the probability distribution for muons with increasing energies between 4 to 24 keV with 4 keV steps. The higher the energy the further the muons penetrate on average into the sample, but this also broadens the profile. Up to 12 keV all muons stop within the N layer and only for higher energies an increasing fraction stops within the S layer.

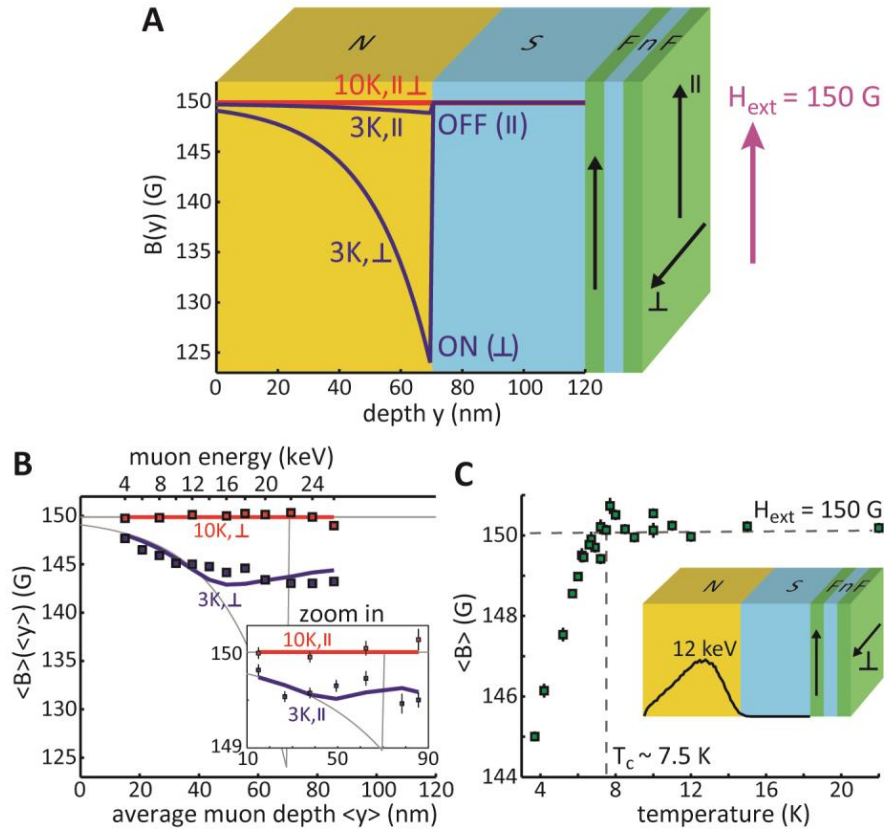

**Fig. 2. Fit results to LE-μSR data on the NSFnF architecture.** (**A**) The magnetic flux profile $B(y)$ obtained from fitting all data simultaneously (at fixed temperature), for both the collinear ($\parallel$) and orthogonal ($\perp$) arrangement. Red for $T = 10$ K and blue for $T = 3$ K. For the latter an exponentially decaying model function was used while the former is taken to be constant. (**B**) The average magnetic flux $\langle B \rangle(\langle y \rangle)$ obtained from fitting each dataset individually (i.e. the conventional treatment) compared to the calculated values from the profiles of (**A**). Top axis shows the corresponding muon energies of the data points. (**C**) Temperature dependence of the average flux $\langle B \rangle$ in the orthogonal arrangement, taken at a muon energy of 12 keV (muon stopping profile displayed in inset) to ensure all muons stopped in the Au layer.

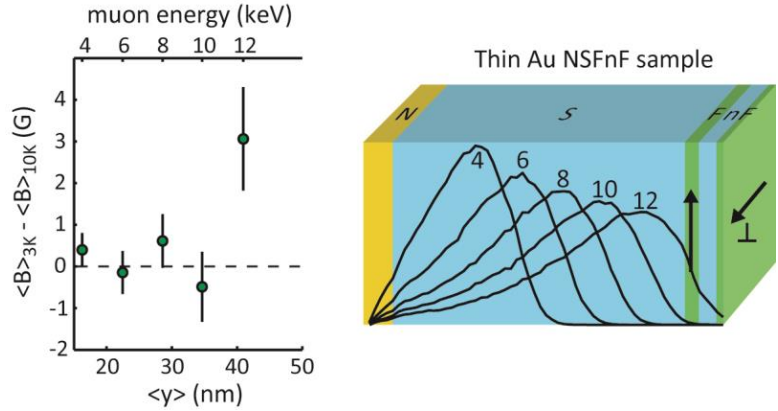

**Fig. 3. Thin Au cap sample.** The difference of the induced magnetic flux at $T = 3$ K and that at $T = 10$ K for the NSFnF architecture with a very thin 5 nm N (Au) cap in the orthogonal arrangement (displayed with muon stopping profiles overlayed on the front face). The highest energy (12 keV) includes contributions from the n-spacer. It is only in the region of the FnF interface that any difference is detected between above and below $T_c$.

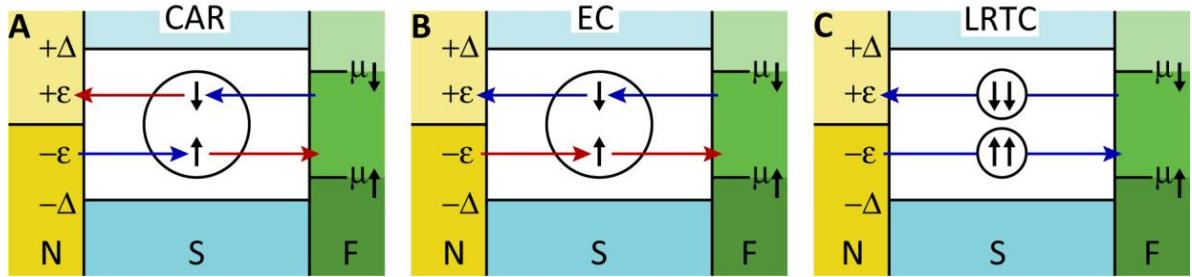

**Fig. 4. Spin-transfer mechanisms.** Schematic of the proposed mechanisms to transfer spin across the superconductor (S) with gap energy $\Delta$ when there is a spin accumulation in the ferromagnet (F) resulting in a shift between the chemical potentials µ of the spin up and spin down band. (**A**) During a crossed Andreev reflection (CAR) a singlet Cooper pair (CP) is created from an electron at energy $+\varepsilon$ with spin down $(+\varepsilon_\downarrow)$ originating from the F layer and an electron at energy $-\varepsilon$ with spin up $(-\varepsilon_\uparrow)$ originating from the normal metal (N) layer (blue arrows). CAR can also annihilate a CP by donating electron $+\varepsilon_\downarrow$ into the N and $-\varepsilon_\uparrow$ into the F layer (red arrows). (**B**) During an elastic co-tunneling (EC) process a singlet CP attracts electron $+\varepsilon_\downarrow$ from the F layer while simultaneously donating its own $+\varepsilon_\downarrow$ electron into the N layer (blue arrows). EC can also attract electron $-\varepsilon_\uparrow$ from the N layer and donate its own $-\varepsilon_\uparrow$ electron into the F layer (red arrows). (**C**) A flow of polarized (triplet) Cooper pairs can transfer spin across the S layer, without generating a moment inside the S layer. Triplet pairs of $+\varepsilon_\downarrow$ electrons move from the F to the N layer while an equal flow of triplet pairs of $-\varepsilon_\uparrow$ electrons move from the N to the F layer.